\documentclass[lettersize,journal]{IEEEtran}
\usepackage{amsmath,amsfonts}
\usepackage{algorithmic}
\usepackage{algorithm}
\usepackage{array}
\usepackage[caption=false,font=normalsize,labelfont=sf,textfont=sf]{subfig}
\usepackage{textcomp}
\usepackage{stfloats}
\usepackage{url}
\usepackage{verbatim}
\usepackage{graphicx}
\usepackage{hyperref}
\usepackage{cite}
\begin{document}

\title{Real-time multichannel deep speech enhancement in hearing aids: Comparing monaural and binaural processing in complex acoustic scenarios}

\author{Nils~L.~Westhausen,
        Hendrik~Kayser,
        Theresa~Jansen,
        Bernd~T.~Meyer%
\thanks{ }
\thanks{Nils L. Westhausen and B. T. Meyer are with the Communication Acoustics group and the Cluster of Excellence "Hearing4all", Carl von Ossietzky Universität Oldenburg, Oldenburg, Germany (e-mail: \{nils.westhausen, bernd.meyer\}@uni-oldenburg.de).}
\thanks{Hendrik Kayser is with the Auditory Signal Processing and Hearing Devices group and the Cluster of Excellence "Hearing4all", Carl von Ossietzky Universität Oldenburg, and the Hörzentrum Oldenburg gGmbH Oldenburg, Germany (e-mail: hendrik.kayser@uni-oldenburg.de).}
\thanks{Theresa~Jansen is with the Hörzentrum Oldenburg gGmbH and the Cluster of Excellence "Hearing4all", Oldenburg, Germany (e-mail: jansen@hz-ol.de).}%
}

\maketitle

\begin{abstract}
Deep learning has the potential to enhance speech signals and increase their intelligibility for users of hearing aids. Deep models suited for real-world application should feature a low computational complexity and low processing delay of only a few milliseconds. In this paper, we explore deep speech enhancement that matches these requirements and contrast monaural and binaural processing algorithms in two complex acoustic scenes. Both algorithms are evaluated with objective metrics and in experiments with hearing-impaired listeners performing a speech-in-noise test. Results are compared to two traditional enhancement strategies, i.e., adaptive differential microphone processing and binaural beamforming. While in diffuse noise, all algorithms perform similarly, the binaural deep learning approach performs best in the presence of spatial interferers. Through a post-analysis, this can be attributed to improvements at low SNRs and to precise spatial filtering. 
\end{abstract}


\section{Introduction}
The improvement of speech
intelligibility for hearing-aid users is a challenging problem in complex acoustic scenes.
Hearing aids often feature multiple microphones on each device, which enables multi-channel processing for the left and right side. 
Signals captured at the relatively densely positioned microphones (with a distance of approximately 1\,cm) at one side can be enhanced, resulting in monaural processing; 
alternatively, true binaural processing can be performed, which requires a transmission of the microphone signals between the two devices. 

One of the most widely adopted algorithms for monaural processing in this context is bilateral adaptive differential microphones (ADM) \cite{Elko1995}. 
The ADM places the region(s) of the least sensitivity in the rear hemisphere of the listener and amplifies sound in the frontal hemisphere to improve the signal-to-noise ratio (SNR). 
The monaural processing of the ADM can result in slight distortion of binaural information \cite{Bogaert2005}.
A traditional approach for binaural processing is the binaural Minimum Variance Distortionless Response beamformer (MVDR) \cite{Marquardt2016} using microphones on both sides of the head, with one instance of the MVDR beamformer running for each side based on the same parameters for directional steering and noise covariance.
While the binaural MVDR is very capable to enhance the target direction, it distorts the binaural cues for sources not radiating from the target direction.

Recently, deep-learning based speech enhancement approaches have shown promising results in the context of hearing aids \cite{Tammen2022, Schröter2022ha, Gajecki2023}. 
Several studies evaluated deep speech enhancement with hearing aids using headphones and audio material that was created offline \cite{Goehring2016,Healy2021, Graetzer2021}. 
Recent studies evaluated signals processed in real time \cite{Diehl2023a, Diehl2023b}, which resulted in promising intelligibility improvements; but they were also limited to single-channel enhancement and did not use the signals of hearing aid microphones. 
A number of additional approaches were suggested for the Clarity Enhancement Challenges \cite{Graetzer2021, Akeroyd2023}. The aim of the clarity challenges is to evaluate machine learning approaches for speech-in-noise conditions in the context of hearing aids. The first and the second enhancement round used simulated hearing aid signals, where in the first only one interferer (localized speech or localized noise) was present, while the second round featured multiple interfering sources. The systems that were successful in
the first challenge were trained for more traditional speech enhancement with a target approximately from the front, while in the second challenge, because of the more complex scenario, the well-performing approaches were trained for target speaker extraction such as in \cite{Cornell2023}. While all proposed systems were causal, the challenges did not aim for efficient models that can be applied in real-time on restricted hardware.

In our earlier work, we proposed an approach that meets two important requirements of hearing devices, i.e., low-latency processing and limited computational resources. 
We introduced  the binaural group communication filter-and-sum network (GCFSnet), which was applied to the problem of speaker \emph{separation} \cite{Westhausen2024}. 
The approach used grouping of latent representations and weight sharing between these groups for a decreased computational complexity while maintaining an algorithmic latency of 2\,ms. 
The model had full access to the bilateral channels of hearing aids without latency and produced dichotic output to preserve spatial cues. 
However, a working implementation of such a bilateral communication link would require a wired connection between the hearing devices, which is not desirable.  

To address this issue, we evaluated the GCFSnet approach in combination with a low-bitrate binaural link between bilateral models \cite{Westhausen2023} and applied it to speech enhancement. 
The models were designed to be compatible with hearing-aid chips used for research (in terms of computational complexity) \cite{Karrenbauer2022}.
While these approaches resulted in promising improvements compared to models without any link and only small performance loss compared to having no delay in the transmission, the evaluations were limited to objective metrics and the experiments were performed with simulated test data from the same domain as the training data; it is therefore unclear how well the algorithms generalize to unseen conditions.
 
In the current study, we explore if speech enhancement with the GCFSnet is beneficial for hearing-impaired, aided listeners with moderate to severe hearing loss in two complex auditory scenes. The measurements were performed in real-time with subjects wearing hearing aid dummies in simulated scenes rendered to loudspeakers.
Another aspect we investigated in this study is if the subjective data can be predicted by objective intelligibility metrics.

The paper is structured as follows: First, the methods used in this study are introduced. These include the theory of deep spatial filtering and post filtering, a description of the architecture and training data, and the model and training configuration. Next, the baseline systems are introduced, followed by the complex acoustic scenes and the hearing-aid configuration. The methods section concludes with a description of the subjective measurements and the subjects, followed by the objective evaluation, including the objective metrics.

In the next section, the results of the subjective and objective evaluations are presented, which is followed the discussion. Finally, the paper concludes with a summary of the main findings.

\section{Methods}

\subsection{Deep spatial and post filtering}
\label{sec:scenario}
In this study, we focus on scenarios with one target speaker in front and additional maskers, which can be either diffuse noise or two interfering speakers. 
Our enhancement approach uses the short-time Fourier transformation (STFT) representation of the mixture $y$ as input, which can be written as 
\begin{equation}
    Y(m, t, f) = X_{S}(m,t,f) + X_{D}(m,t,f). 
\end{equation}
In this notation, $Y$ denotes the STFT of the mixture where as $m$, $t$, $f$ are the microphone, frame and frequency index, respectively.
$X_S$ is the reverberant STFT of the target speech $s$ convolved with the impulse response of the target speaker $h_s$.
$X_D$ corresponds to the STFT of the interfering sources $d$. 

\begin{figure}
  \centering
  \includegraphics[width=1.00\linewidth]{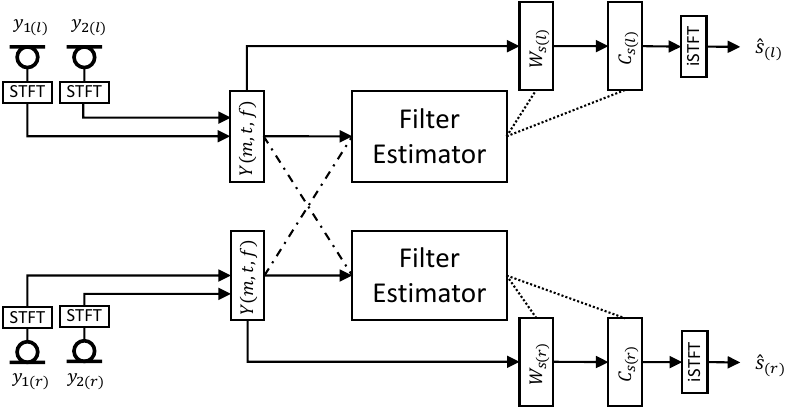}
  \caption[Illustration of the proposed approach for spatial filtering and post filtering.]{Illustration of the proposed approach for spatial filtering and post filtering. The filters for the left and right side are estimated by separate models. The dashed-dotted line symbolizes an optional exchange of the complex TF representation of microphone signals for binaural input features.}
\label{fig:filter-structure}
\end{figure}

We aim to extract the source estimate $\hat{s}$ including the direct part as well as early reflections. 
The extraction is performed for the left and right side separately with complex filter-and-sum beamforming
\begin{equation}
\label{eqn:filter_channel}
\Tilde{S}(t, f) = \sum_{m=1}^{M} Y(m, t, f) \cdot W(m, t, f).
\end{equation}
$\Tilde{S}$ denotes an intermediate time-frequency (TF) representation of the estimated target speaker and $W \in \{ z \in \mathbb{C} \mid -r \leq \Re(z) \leq r, -r \leq \Im(z) \leq r \}$ corresponds to the complex filter weights estimated by the model where $r$ is the learned range of the real and imaginary components. 
$M$ is the number of microphones used for the filtering. 
In the next step, a single frame postfilter is applied with the aim to further improve the signal
\begin{equation}
\label{eqn:filter_post}
\hat{S}(t, f) =  \Tilde{S}(t, f) \cdot C(t,f).
\end{equation}
$C \in \{ z \in \mathbb{C} \mid -r \leq \Re(z) \leq r, -r \leq \Im(z) \leq r \}$ are the complex filter weights, where the numeric range of filter coefficients $r$ is a learnable parameter. $\hat{S}$ is transformed back to the time domain by an inverse STFT.
The general structure of the filtering is illustrated in \autoref{fig:filter-structure}.
\subsection{Architecture}
\begin{figure*}
  \centering
  \includegraphics[width=1.00\linewidth]{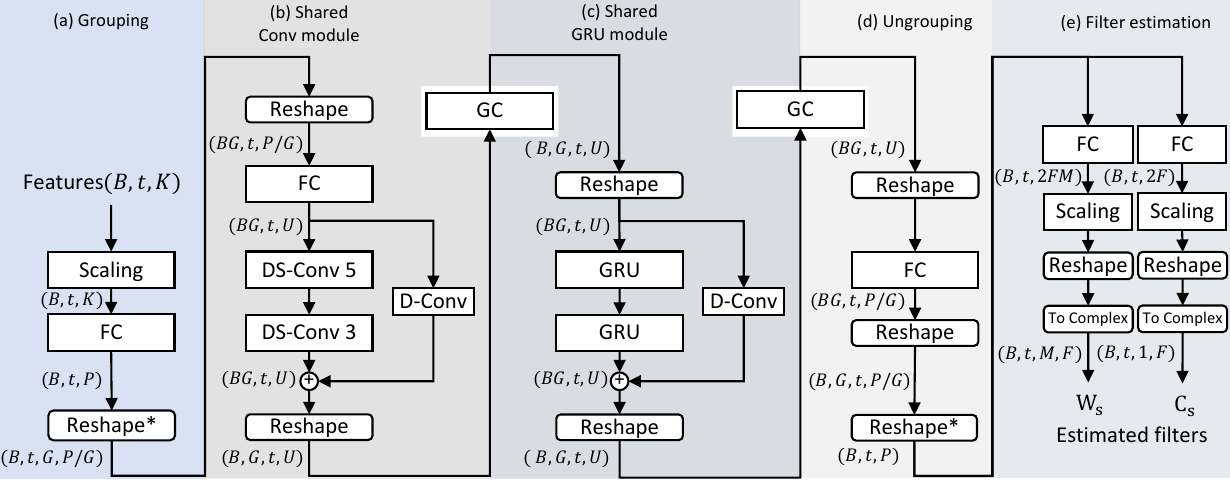}
  \caption[Illustration of the proposed filter estimation model.]{Illustration of the proposed filter estimation model. Reshape operations with (*) include axes permutation.}
\label{fig:filter-network}
\end{figure*}

In this study, speech enhancement is performed with the \emph{group communication filter and sum network} (GCFSnet) introduced in \cite{Westhausen2023}. In \cite{Westhausen2023}, the model was introduced as an architecture with low computational and memory footprint enabled by grouping of latent representations and weight sharing between groups.
The architecture is illustrated in \autoref{fig:filter-network}; it contains five modules and uses group communication \cite{Luo2021gc} and weight sharing for a reduced computational footprint. 
First, the feature vector of size $B$ is scaled by a single learned parameter in the grouping module. This scaling is performed to learn the optimal input range to the model.
This feature vector is then projected by a fully connected (FC) layer with shape $B \times P$ and a tanh activation function. This projection mixes the feature vector and distributes the information to specific groups.
The latent representation of size $P$ is split in $G$ equal groups. 

The second step is represented by the conv module which is shared between groups.
In this module, an FC layer with tanh activation and dimension $(P/G)\times U$ maps the latent representation of the group to the hidden size $U$. 
The representation of size $U$ is processed by two causal depth-wise separable convolution layers (DS-Conv), the first with kernel size 5 and the second with kernel size 3 in time dimension. DS-Conv is utilized to save parameters and operations compared to a standard 1D-Conv layer.
Inspired by \cite{Schröter2022filternet2}, a skip connection with a depthwise convolution (D-Conv) with kernel size 1 for scaling is implemented in parallel to the DS-convs.
The idea of the conv module is to capture information on a shorter time scale with the help of the fixed causal contexts of the DS-Conv operations.  

The next step is a group communication (GC) block. GC enables the model share information between the parallel groups. In our own previous work (\cite{Westhausen2023} and \cite{Westhausen2024}) GC was implemented with transform average concatenate (TAC) as suggested in \cite{Luo2021gc3}.
For the current study, we exchanged TAC with a simpler structure we refer to as group communication with group mixing:
First, a FC layer of shape $U \times (P/G)$ with tanh activation maps the hidden representations of the groups to size $(P/G)$.
The mapped representations are concatenated to form a latent representation of size $P$.
This representation is mixed by an $P \times P$ FC layer with tanh activation.
Next, this intermediate representation is split in $G$ groups and mapped by an FC layer with shape $(P/G)\times U$ and tanh activation. 
In the final step of this block, the input to the mixing module is added to the output as a skip connection. 
This group mixing mechanism is slightly less complex compared to the TAC with respect to implementation and is restricted to fixed-range intermediate representations (due to the use of the tanh activation), which can be advantageous for fixed-point implementations. 

The third module is the shared GRU (gated recurrent units \cite{Cho2014}) module. This module includes two consecutive GRU layers with $U$ units and an additive skip connection with a D-Conv for scaling.
In 
Because of its recurrent nature and gating mechanisms the GRU module is intended to capture more long-term temporal information. In \cite{Cho2014} GRUs have shown these capabilities for statistical machine translation. 

GC is again applied after the GRU module.
Next, the hidden representations of the groups are mapped to size $P/G$ by an FC layer of size $U \times (P/G)$. 
The groups are combined to form a latent representation of size $P$. 
In the final step of this module, the real and imaginary parts of the filter weights $W$ and $C$ are predicted by FC layers with tanh activations and scaled by the learned parameter $r$ as mentioned in \autoref{sec:scenario}. 
\subsection{Training data generation}
Robust speech enhancement models require extensive data for training.
In this study, we generated fixed training and development sets.
The training set comprised 160k samples, while the development set included 3k samples, each 4 seconds long with a 16\,kHz sampling frequency.
Speech data was sourced from the DNS-Challenge dataset \cite{Reddy2020}, and noise data from three datasets: DNS-Challenge \cite{Reddy2020}, the first Clarity enhancement challenge dataset \cite{Graetzer2021}, and LibriMix \cite{Cosentino2020}.
We also simulated 60k multichannel binaural room impulse responses (BRIR) using RAZR \cite{Wendt2014}.
To enable the simulation of hearing-aid microphone signals for the hearing aid dummies of the PHL we measured an HRTF set with an angular resolution of 1° using a KEMAR dummy head.
The setup for this measurement is described in \cite{Blau2018}.
The limitation of the loudspeakers in this setup made a low-frequency extension necessary which is explained in \cite{Stärz2022}. 
This set features the four hearing-aid microphones with the devices placed on the KEMAR artificial head, front-left, front-right, left-back and right-back.

Our simulated rooms varied in size, with widths ranging from 3 to 10\,m, heights between 2.5 and 4\,m, and surface areas from 12 to 100\,$\mathrm{m}^2$. 
The reverberation time (T60) ranged from 0.25 to 1\,s.
The receiver was positioned no more than one meter from the center of the room at a height of 1 to 1.4\,m.
Each scene featured a target speaker placed within an azimuth range of -10° to 10° azimuth.
Two interfering speakers were placed at different angles while omitting the -20° to 20° azimuth range. 
A minimum angular separation of 10° between the speakers was enforced.
The speakers' distance ranged from 0.75 to 2m, at heights of 1 to 1.4\,m.
Additionally, two noise sources were simulated: one localized interferer from the DNS-Challenge dataset or the Clarity dataset, and another for diffuse-like noise, created using the RAZR feedback delay network output combined with ambient noise from LibriMix, which is based on the WHAM! corpus \cite{Wichern2019}. Both noise sources were positioned at least 1\,m away from the receiver, at heights of 1 to 1.4\,m.

For each interfering speaker, a random better-ear SNR between -8 and 8\,dB was sampled from an equal distribution.
The noise sources were mixed at a relative SNR sampled from $\mathcal{N}(0, 5^2)$\,dB.
The full noise signal is then mixed with target speech with a better ear SNR equally distributed between -8 and 8\,dB.
The mix is scaled to a value drawn from $\mathcal{N}(-28, 10^2)$\,dB relative to full scale.
One part of the training and development set contained only two interfering speakers (30\% of the scenes), another part contained only interfering noise (30\%), and another part contained both speech and noise interferers (40\%).

For the training target the BRIR of the front microphones for the target speaker was windowed to include the direct part of the BRIR as well as the some early reflections up to 300~ms. BRIRs are windowed as described in \cite{Braun2021}.
To obtain the final binaural target the target speech signal was convolved with the windowed BRIR.

\subsection{Model and training configuration}
To quantify the contribution of binaural communication between hearing aids, we compared two versions of the network, both of which produced dichotic output which is either based on binaural features (\emph{GCFSnet(b)}) or on bilateral processing that only relied on monaural features (\emph{GCFSnet(m)}).
The concatenated real and imaginary part of the frequency representation served as input features as suggested in \cite{Westhausen2024}. 
For the binaural version, we assumed a transmission without delay between the ears.
The models on both sides shared the same weights, while the channel order in the binaural case (\emph{GCFSnet(b)}) for each side was front microphone on the ipsilateral side (1), front microphone on the contralateral side (2), back microphone on the ipsilateral side (3) and back microphone on the contralateral side(4). For \emph{GCFSnet(m)} the channel order is the front microphone on the ipsilateral side (1) followed by the back microphone on the ipsilateral side(2).
Following \cite{Westhausen2023}, $P$ was set to 128 and $G$ was set to 8, and $U$ was set to 32. 
This resulted in 135k weights for the monaural version and in 168k weights for the binaural version. 
The model sizes were chosen based on the constraints of the research hearing aid chip described in \cite{Karrenbauer2022}.
The models were trained with quantization-aware training \cite{Jacob2018} for weights and biases. 
The weights were linearly quantized to 8 bits and the biases to 16 bits with both having a range between -1 and 1.
This reduced the size of the model drastically compared to keeping all weights and biases at 32 bits. 
The calculations inside the model were not quantized.
The frame length was 4\,ms and the frame shift was 2\,ms. 
The FFT size was 128 with equal front and back padding. 
A hann window was applied for the forward transformation. 
These parameters were given by the signal processing chain of the openMHA setup used in this study. 

The model was trained with compressed spectral Mean Squarred Error (cMSE) as proposed in \cite{Braun2021loss}. 
The STFT configuration of the loss function is independent of the one used for the filtering framework since it was calculated after signal reconstruction. The STFT of the loss function was set a to window length of 20\,ms with a 10\,ms shift and an FFT size of 320. 
The model was trained for 50 epochs with an initial learning rate of 2e-3 and the ADAM optimizer in Tensorflow 2.11. 
In each epoch, the learning rate was multiplied by 0.98. 
If the loss on the development set did not decrease for 5 consecutive epochs, the loss was multiplied by 0.5.
For gradient clipping, AutoClip \cite{Seetharaman2020} was applied with $p=10$ for smoother training and better generalization. 
The weights were saved when the loss on the development set decreased. 
For running the models inside the openMHA, the full architecture was implemented in standard lib C without any special optimizations.

\subsection{Baseline algorithms}
The two configurations of the GCFSnet were compared with two traditional algorithms commonly used in hearing aids.
The first was the adaptive differential microphone (ADM) \cite{Elko1995} which uses the two microphones in one hearing aid to create a monaural filter.
The ADM was configured to optimize its output for placing the region of the least sensitivity (Null) in the rear hemisphere.
ADM algorithms on the left and right side work independently, which can slightly distort the binaural perception.

The second was a fixed binaural Minimum Variance Distortionless Response beamformer (MVDR) \cite{Marquardt2016} steered to 0° azimuth. 
This beamformer uses all four microphones available in the hearing aid setup without delay for the transmission. 
It aims at minimizing the power of the output signal with the constraint that a signal arriving from the front is preserved under the assumption of an isotropic, spatially diffuse noise field. 
It produces an output signal that only contains binaural cues related to its target direction, i.e., binaural cues related to sound sources located at other directions are not preserved. 

\subsection{Complex acoustic scenes}
The measurements were conducted in two complex acoustic scenes rendered to a ring of 16 GENELEC 8030 loudspeakers with a radius of 1.56\,m and a height of each loudspeaker of 1.18\,m. The azimuth angle between the loudspeakers was 22.5°.
The loudspeakers were connected to a RME ADI-8 Pro soundcard with a 44.1\,kHz sampling frequency. 
The room had the dimensions of 5.93\,m × 5.01\,m × 2.74\,m  (length × width × height) and a low reverberation time of around 170\,ms.
The simulation was performed with TASCAR \cite{Grimm2019tascar}. The scenes were rendered at a sampling frequency of 44.1 kHz.
The first scene contained a target speaker from 0° azimuth and two localized interfering speakers from +60° and -60° azimuth (\emph{S0N$\pm$60 IFFM}). 
Utterances from the  Oldenburg sentence test with a female talker were used as target sentences, while the interfering signals were based on the International Female Fluctuating Masker (IFFM), a version of the International Speech Test Signal (ISTS) with shortened gaps \cite{Holube2010}. 
The ISTS is an non-intelligible speech signal built from segments of Arabic, English, Mandarin, Spanish, German and French, which enables reproducible measurements. 
Interfering signals were continuously presented during each measurement trial at 65 dB SPL (sound pressure level). 
The level of the target was varied during the measurement to adjust the SNR.
The second scene contained a target speaker from 0° azimuth and diffuse cafeteria noise (\emph{S0Ndiff Cafeteria}).
The cafeteria noise was taken from \cite{Grimm2019recording}.
The noise is a first-order ambisonics recording performed at lunch time in the cafeteria at the University of Oldenburg.
All intelligible speech segments were removed in post processing. 
As for the first scene, the continuous noise was presented at 65\,dB\,SPL while the target speech level was varied.
The subjective measurements in these scenes should be informative if trained deep algorithms could generalize to scenes that include head movements or effects from individual physical traits such as head geometries.

\subsection{Hearing-aid configuration}
Hearing-aid processing was simulated with the open Master Hearing Aid (openMHA) \cite{Kayser2022} software running on a laptop CPU. 
Signals were recorded with hearing-aid dummies, each with two MEMS microphones, of the portable hearing lab (PHL) \cite{Pavlovic2019} worn by the listeners. 
With this setup, effects of individual heads were taken into account.
A MOTU A8 sound card was used to capture the hearing aid input signals at a sampling frequency of 48\,kHz. 
The microphones were calibrated with diffuse speech-shaped noise. 

For signal enhancement, the recorded signals were downsampled to 16\,kHz in the openMHA. 
The bilateral ADM processed the signals in the time domain, while the MVDR beamformer and the GCFSnet configurations operate in the frequency domain. 
The frame length of the hearing-aid processing was 2\,ms,
the window length was 4\,ms and the FFT length was 8\,ms corresponding to 128 samples. 
The enhancement algorithms can be bypassed for the condition \emph{unprocessed}. 
Next, a multi-band dynamic range compressor (MBDRC) was applied for gain control and compression. 
The MBDRC was followed by a frequency shifter for feedback reduction. 
Next, an equalization was performed to compensate spectral effects caused by the closed coupling of the hearing aid receivers in the ear canal. Finally, the signal was resampled to 48\,kHz and presented over the hearing-aid receivers.

Personal ear molds were manufactured for each subject for a closed hearing-aid fitting with receivers in the ear canal. 
The closed fitting reduces feedback and comb-filter effects and allows a higher gain in the low frequencies compared to an open fitting. 
Subjects were fitted with a loudness-based gain prescription rule \cite{Oetting2018}. 
The maximum allowed gain was set to 30\,dB and the maximum output was limited at 100\,dB.
The measured algorithmic latency of the hearing aid setup was 5.4\,ms and the latency of the audio hardware and server was 8.7\,ms. This results in a total latency of 14.1\,ms for the whole hearing aid setup.

\subsection{Subjective measurement procedure}
For evaluating the intelligibility of the enhancement algorithms, speech reception thresholds (SRT) were measured with the Oldenburg matrix sentence test (OLSA) \cite{Wagener1999, Wagener2005}. The OLSA is a procedure where the SNR is changed adaptively during the measurement based on the subject's responses. 
The SRT is calculated by a least-squares fit.
The speech material is structured in lists of 20 sentences where each list is used for one measurement trial.
All sentences have the same structure which is [Name], [Verb], [Numeral], [Adjective], [Noun], with 10 alternatives per category.
The subjects used a touch screen with the 5x10 alternatives to log their response.

In this study, the adaptation procedure is configured to measure the SRT80, i.e., the SNR with 80\% of the words being correctly identified 
which results in SNRs more similar to SNRs encountered in real-world scenarios as when measuring the traditional 50\% correct point \cite{smeds2015}.

The measurements were performed for 6 different conditions, i.e., without any signal modifications  (\emph{Unaided}), with prescribed gain (\emph{gain only}), and the four enhancement algorithms (\emph{ADM}, \emph{MVDR}, \emph{GCFSnet(m)}, \emph{GCFSnet(b)}) each with the prescribed gain. 
For each subject, the SRT80 was measured twice for each condition, resulting in test and retest scores.

\subsection{Subjects}
\begin{figure}
  \centering
  \includegraphics[width=1.00\linewidth]{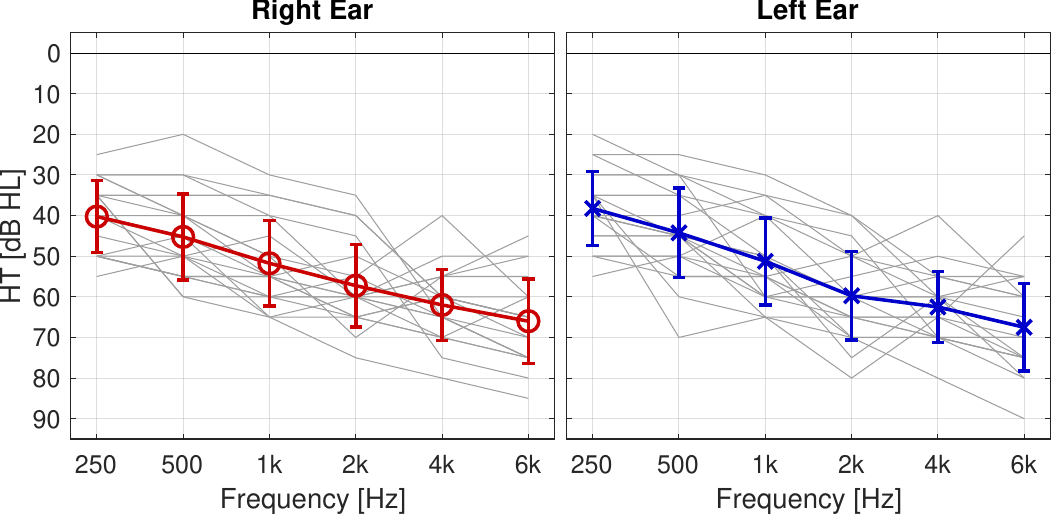}
  \caption[Hearing thresholds of all subjects.]{Hearing thresholds (HT) in dB hearing level (HL) of all subjects. The colored lines show the mean HT and the standard deviation. Individual audiograms are shown in gray.}
\label{fig:ht}
\end{figure}

The evaluation of the algorithms was performed with 20 hearing-impaired subjects (age: 49-81 years, 12 female listeners).
All subjects had a symmetric hearing loss corresponding to N3 or N4 after the Bisgaard standard audiograms \cite{Bisgaard2010}.
N3 corresponds to sloping, moderate hearing loss and N4 to a sloping, moderate to severe hearing loss.
All subjects were native German speakers and were fitted with hearing aids for more than 24 months.
They gave their written consent prior to inclusion in the study.
The experiment was approved by the ethics committee (“Kommission für Forschungsfolgenabschätzung und Ethik”) of the Carl von Ossietzky Universität in Oldenburg, Germany (Drs.EK/2021/031-05).
The mean hearing loss and standard deviation over all subjects is shown in \autoref{fig:ht}.

\subsection{Objective evaluation}
In addition to the subjective evaluation with hearing-impaired subjects, an objective evaluation was performed based on two metrics. The first is the better ear hearing aid speech perception index with a mapping function related to a keyword recognition task (HASPI w2) \cite{Kates2023}.
The second is a combination of the MSBG hearing loss model \cite{Nejime1997} and the modified binaural short time objective intelligibility (MBSTOI) \cite{Andersen2018} as previously used in the first round of the first Clarity enhancement challenge \cite{Graetzer2021}. 

These two metrics were chosen since they relate to speech intelligibility prediction and allow the consideration of individual hearing loss. 
HASPI was specifically built to evaluate hearing aid algorithms, while MBSTOI explicitly takes  binaural perception into account.

Both metrics require clean reference signals without the additive noise and simulated reverberation.
These signals are not part of the subjective measurements and were therefore simulated. 
To this end, a binaural synthesis was conducted using TASCAR to simulate the acoustic scenes.
This synthesis was performed using HRTFs of the hearing aid dummies on a KEMAR artificial head measured inside the loudspeaker ring used during the subjective measurements. 
The simulation enabled us to create separate, time-aligned reference signals for the objective metrics. 
To obtain the reference signal, we deactivated the reflecting surfaces, the image source model and the feedback delay network of TASCAR, so the reference contains no reverberation.

We simulated the signals at fixed SNRs.
The metrics were averaged between -5 and 10~dB SNR (with 1~dB step) since this is an SNR range that covers challenging listening environments in which hearing-aid users would most likely profit from improvements. 
At each SNR, 20 OLSA sentences were synthesized at 44.1kHz. 
Each sentence had a random noise sample drawn from the original noise signals. 
The mixed signals were resampled to 48~kHz to match the input sampling frequency of the hearing aid processing.
All simulated signals were processed with the openMHA setup individually for each subject including hearing loss compensation and enhancement conditions but with deactivated equalization for the hearing aid receivers since the signals are not played back. The results are averaged over all 20 sentences for each SNR.

Additionally, we evaluated the attenuation of the enhancement algorithms depending on the incidence angle.
This evaluation was based on the TASCAR setup for the objective evaluations described above. 
We rendered 20 OLSA sentences for incidence angles from -180 to 180 in 5 degree steps. 
The source had a level equivalent to 70 dB SPL. The simulated room was not changed besides deactivating the noise sources so that the effects of room reverberation are taken into account. 
The attenuation was calculated as the ratio between the energy of the signals processed by the enhancement algorithms and signals processed without activated algorithms. 
The result was averaged over all 20 sentences.

\section{Results}
\begin{figure*}
  \centering
  \includegraphics[width=1.00\linewidth]{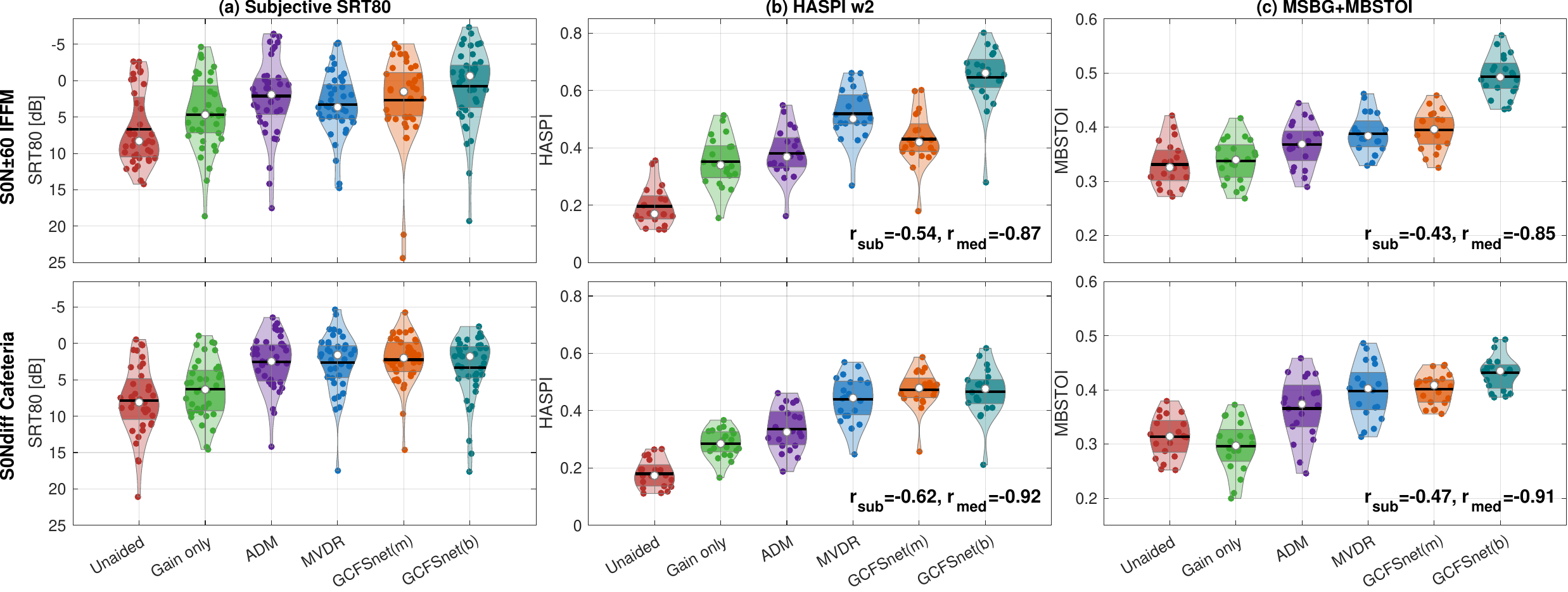}
  \caption[Results for different HA enhancement strategies.]{Results for different HA enhancement strategies for subjective listening tests (left column) and objective metrics HASPI (middle column) and MBSTOI (right column). The mean is marked by black bars and the median by white circles inside the violin. The axes are reversed for the left column. Violin plots are shown for two acoustic scenes (top and bottom row, respectively). The $r$-values show the correlation of the objective metric with the subjective measurements, both on subject level ($r_{sub}$) and for the medians ($r_{med}$). }
 
\label{fig:results}
\end{figure*}

\autoref{fig:results} shows violin plots of the results of the subjective SRT measurements for all subjects, conditions and scenes together with the objective evaluation for the same subjects performed with HASPI w2 and MSBG+MBSTOI.

According to a Kolmogorov–Smirnov test, all 36 data distributions (6 (algorithms) $\times$ 3 (metrics) $\times$ 2 (room configurations)) are normally distributed (p $\leq0.05$). Seemingly, two violin plots in \autoref{fig:results} (for instance, \emph{ADM} in the \emph{S0N$\pm$60 IFFM} Scene with subjects) suggest the data to follow a bimodal distribution, which is a result from the local density estimator of the violin plot algorithm (Hintze and Nelson, 1998 \cite{Hintze1998}) and does not reflect the outcome of the statistical test. To analyze the statistical differences between means, we conducted a one-way ANOVA of the five algorithms that actually change the signal, i.e., the condition \emph{unaided} - which performs worst in five out of six conditions - was excluded from the analysis. For each condition, the ANOVA indicated significant differences. We therefore conducted a t-test between the top two algorithms, which indicated significant differences in four out of the six conditions. An overview of the p-values associated with the ANOVA and the subsequent t-tests, as well as the best algorithms and their means is provided in \autoref{tab:stat}.

\begin{table*}
\caption{Results of the statistical analysis: For each scene, the table reports the p-value associated with a one-way ANOVA for five processing algorithms (excluding "unprocessed" \emph{Unaided}). The following columns report the best and second-best algorithm, the corresponding mean scores in parentheses and the t-test p-value of their comparison. p-values below the significance level $p = 0.05$ are marked with an asterisk. For the subjective scores, lower values are better, while for HASPI and MBSTOI, higher values are better. The subjective SRT80 is reported in $\mathrm{dB}$.}
    \centering
    \begin{tabular}{lccccc}
                           & \textbf{Scene}      & \textbf{ANOVA}    & \textbf{Best}               & \textbf{2nd best}           & \textbf{t-test}\\
                           &            & p-value  &                    &                    & p-value\\
    \hline
         Subjective SRT80 & S0N$\pm$60 & 0.024 \textbf{*}    & GCFSnet(b) (0.759) & ADM (2.1)          & 0.0224 \textbf{*}\\
                        & S0Ndiff    & $3.10 \times 10^{-5}$ \textbf{*} & GCFSnet(m) (2.23)  & ADM (2.51)         & 0.531\\
    \hline
         HASPI             & S0N$\pm$60 & $2.40 \times 10^{-16}$ \textbf{*} & GCFSnet(b) (0.647) & MVDR (0.519)       & $9.65 \times 10^{-9}$ \textbf{*}\\
                           & S0Ndiff    & $1.60 \times 10^{-14}$ \textbf{*} & GCFSnet(m) (0.473) & GCFSnet(b) (0.466) & 0.367\\
    \hline
         MBSTOI            & S0N$\pm$60 & $1.80 \times 10^{-21}$ \textbf{*} & GCFSnet(b) (0.493) & GCFSnet(m) (0.395) & $4.14 \times 10^{-21}$ \textbf{*}\\
                           & S0Ndiff    & $6.40 \times 10^{-15}$ \textbf{*} & GCFSnet(b) (0.432) & GCFSnet(m) (0.402) & $1.82 \times 10^{-12}$ \textbf{*}\\
    \hline
    \end{tabular}
    
    \label{tab:stat}
\end{table*}
\subsection{Results of subjective measurement of speech intelligibility}
The results of the listening experiment for the first acoustic scene with spatial maskers (IFFM), the ANOVA indicated a significant difference of means (p $\leq0.05$) of the processing algorithms compared to each other.
From the two best-performing algorithms (\emph{ADM}) and \emph{GCFSnet(b)}), \emph{GCFSnet(b)} performs significantly better (with a mean SRT80 0.8) according to a paired t-test ($p \leq 0.05$) and achieves a mean SRT that is 1.3~dB lower (better) than \emph{ADM}. 
The SRT improvement of \emph{GCFSnet(b)} over \emph{Unaided} and \emph{Gain only} is 5.9~dB and 3.9~dB, respectively.  

In the diffuse noise scenario, a one-way ANOVA indicated significant differences between the algorithm-specific means ($p \leq 0.001$).
Among these, \emph{GCFSnet(m)} achieved the most favorable mean SRT at 2.2 dB, followed by \emph{ADM} with an SRT of 2.5 dB. The t-test did not identify a significant difference between these two algorithms ($p > 0.05$). Notably, all enhancement algorithms demonstrated improvements in comparison to the \emph{Unaided} condition, which had an mean SRT of 7.9 dB, and the \emph{Gain only} condition, with an SRT of 6.3 dB.

\subsection{Results in terms of HASPI}
The HASPI w2 results are shown in column (c) of \autoref{fig:results}.
A one-way ANOVA of the HASPI scores indicated a significant difference among the algorithms in both acoustic scenes ($p \leq 0.001$ in both cases). 
In the \emph{S0N$\pm$60 IFFM} scene, similar to the subjective results, \emph{GCFSnet(b)} achieved the best performance with the highest mean HASPI score of 0.65.
Its performance was significantly better than the second-best algorithm \emph{MVDR} with a HASPI score of 0.52 (based on a paired t-test, $p \leq 0.001$).
For \emph{S0Ndiff Cafeteria}, \emph{GCFSnet(m)} and \emph{GCFSnet(b)} recorded the highest mean HASPI scores of 0.47. The t-test did not show a significant difference for the these algorithms ($p > 0.05$). 

\subsection{Overall Results in terms of MBSTOI}
The MBSTOI results are shown in column (c) of \autoref{fig:results}. 
We performed the same statistical analysis as described above, which indicated significant differences between processing conditions ($p \leq 0.001$) for both acoustic scenes.
In both acoustic scenes, \emph{GCFSnet(b)} achieved the highest means (0.49 and 0.43 for the IFFM and Cafeteria scene, respectively) which in both cases was significantly better than the runner-up \emph{GCFSnet(m)} (paired t-test, $p \leq 0.001$). 
For the Cafeteria scene, this is different from the subjective results, where \emph{ADM} performed best and was not statistically different from the best-performing algorithms. 

\subsection{Correlation between subjective and objective results}
To quantify how well objective metrics can predict responses from aided, hearing-impaired listeners, we analyze the Pearson correlation between the SRT80 from listening experiments and the according score obtained from HASPI and MBSTOI.
We report the correlation $r_{sub}$ that takes into account data from individual subjects in all conditions (i.e., all colored data points in \autoref{fig:results}) as well as the correlation $r_{med}$ based on median values in which the subject-level scores are pooled. 
The corresponding scores are shown in the lower right corners of the HASPI and MBSTOI panels in columns (b) and (c). 
All correlation values are significant at $p \leq 0.05$.
Negative correlation coefficients are obtained since a better subjective performance results in a \emph{lower} SRT80 value; note that the scales in the left column of \autoref{fig:results} are inverted.
The correlations for individual subjects are moderate, while very strong correlations are obtained when the medians are considered. 

\subsection{Results metrics depending on SNR}
\begin{figure}
  \centering
  \includegraphics[width=1.00\linewidth]{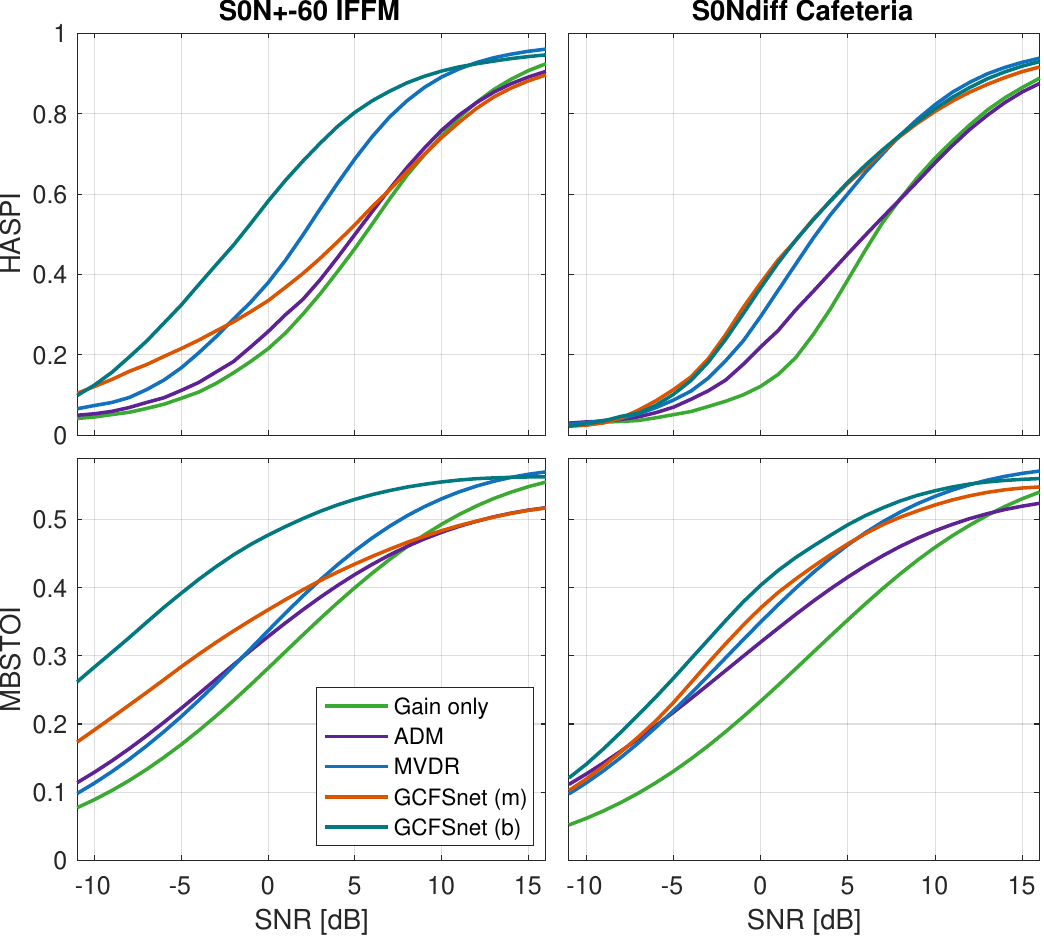}
  \caption[Objective metrics versus SNR.]{Objective metrics in terms of HASPI (first row) and MBSTOI (second row) pooled over all subjects plotted against SNR for both acoustic scenes.}
\label{fig:metric_snr}
\end{figure}

In a post analysis, we explored the objective metrics depending on the SNR to gain insight how the enhancement algorithms perform in different noise conditions. 
\autoref{fig:metric_snr} shows the HASPI and MBSTOI scores plotted against the SNR for each scene individually. 
The results are pooled over all listeners, which are considered through their individual hearing loss that is used as input to the metrics.
For \emph{S0N$\pm$60 IFFM}, \emph{GCFSnet(b)} produces a benefit in an SNR range of -5 to 5, for both HASPI and MBSTOI. 
For high SNRs, the metrics saturate and differences become very small. 
At low SNRs, \emph{GCFSnet(m)} is predicted as second-best algorithm, but is outperformed at higher SNRs. 
The \emph{MVDR} shows a steeper performance increase with SNR than \emph{ADM} or \emph{GCFSnet(m)} in both metrics. 
In the \emph{S0Ndiff Cafeteria} conditions, the objective scores for \emph{GCFSnet(b)}, \emph{GCFSnet(m)} and the \emph{MVDR} are very similar, i.e., both HASPI and MBSTOI do not predict SNR-specific gains in this diffuse-noise condition.
\emph{ADM} and \emph{Gain only} produce lower objective scores, with a slight benefit of \emph{ADM} over \emph{Gain only} at low SNRs. 

\subsection{Results for attenuation over incidence angle}
\begin{figure}
  \centering
  \includegraphics[width=1.00\linewidth]{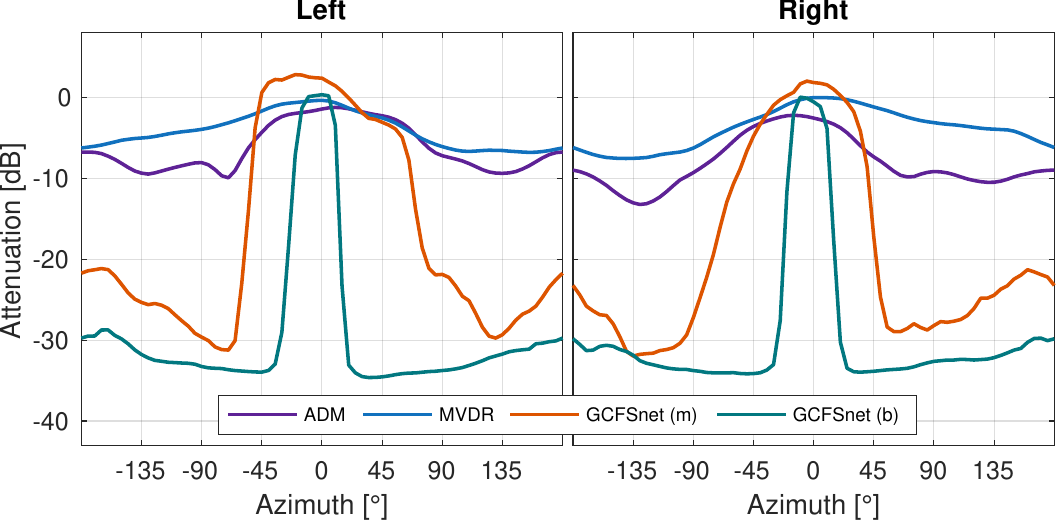}
  \caption[Attenuation over incidence angle.]{Attenuation of clean speech over incidence angle for the left and the right hearing aid device separately.}
\label{fig:angle_att}
\end{figure}

Since all algorithms perform multi-channel processing and spatial filtering, we analyze the attenuation depending on the azimuth angle. 
The results are shown in \autoref{fig:angle_att}. 

The attenuation patterns of the algorithms are quite different: 
\emph{GCFSnet(b)} exhibits the narrowest beam combined with strong attenuation: The full attenuation of -30 dB is already reached at $\pm$45°. 
\emph{GCFSnet(m)} also produces strong attenuation for angles deviating from 0° but with a clearly broader beam. 
An angle-dependent attenuation is also observed for both \emph{ADM} and \emph{MVDR}, but the overall attenuation is lower (with a lowest value of -13~dB). 
The beam patterns are slightly asymmetric, which could results from the measured HRTFs or the simulation of the room. 

\section{Discussion}
\subsection{Comparison of performance of enhancement strategies}
For the \emph{S0N$\pm$60 IFFM} scene, the subjective and the objective metrics show a significant benefit for \emph{GCFSnet(b)}, suggesting that the binaural deep learning approach is a good choice in the presence of localized speech interferers. 
A possible contribution to this result is the narrow beam associated with \emph{GCFSnet(b)} as depicted in \autoref{fig:angle_att}. 
The maximum attenuation is already reached at the location of the interferers at $\pm$60°. 
The binaural features enable the model to accurately identify the direction of a speech source and to attenuate other sources outside the azimuth range defined during the training. 
Similarly, \autoref{fig:metric_snr} shows a clear benefit over large SNR range for \emph{GCFSnet(b)}.
In contrast, \emph{GCFSnet(m)} does not produce benefits compared to the traditional methods for the \emph{S0N$\pm$60 IFFM} scene. 
A potential reason becomes apparent in \autoref{fig:angle_att}: 
The monaural model produces an attenuation pattern wider than its binaural counterpart and does not reach the same level of attenuation, i.e., the suppression of interfering speakers will be less effective.
However, the monaural approach still reaches attenuation levels of 20\,dB for clean speech 
or more although it is limited to the signals of two closely-spaced microphones. 
In the scene with diffuse noise, we did not observe significant differences between traditional and neural systems, except for the gain-only condition, which was outperformed by the other algorithms. 
It is plausible that the narrow beam of \emph{GCFSnet(b)} does not provide strong benefits in conditions with diffuse noise only.

ADM produces good subjective results in both acoustic scenes, which could be linked to the partial preservation of binaural cues and a processing strategy that does not introduce stronger speech distortions.
The performance of ADM is underestimated by the objective measures, especially for HASPI. This could be linked to the fact that HASPI does not consider binaural effects apart from better-ear listening and both metrics could not consider learning effects. 
Several subjects reported that the deep-learning based processing sounded unusual, which could indicate that GCFSnet algorithms produce artifacts different from ADM and MVDR. We presume that a longer exposure to GCFSnet processing could help listeners to adapt to these artifacts, potentially increasing performance with these algorithms. 

\subsection{Relation between objective and subjective results}
The results of the objective evaluation of SRT80 and the evaluation in terms of HASPI and MBSTOI following a similar pattern as shown in \autoref{fig:results}. 
On the individual level, only moderate correlations are observed (with $r$ ranging from -0.43 to -0.62). 
We assume this can be partially attributed to variability caused by factors related to hearing impairment beyond the audiogram, or fatigue and the cognitive condition, which are not accessible to the models. 
When this variability is partially removed by calculating the median performance, very strong correlations of 0.85 and higher are obtained, which suggests that the objective metrics are a good indicator for the enhancement benefit when data is pooled over subjects. 
We find this encouraging since the objective measures used a binaural, synthetic input generated with TASCAR at a fixed SNR set, and head differences or head movements have not been considered.

\subsection{Generalization between training, measurement and simulation setup}
Some properties of the training data were similar to the testing condition (e.g., the subset with interfering speakers always contained two speakers) while most parameters differed (by using random random rooms, SNRs, source and receiver positions, including their height) and others were chosen to be disjunct from the test set (speaker identity and consequently signals, interferer signals). 
The training data only contained static scenes based on BRIR simulations combined with a single HRTF set of the hearing-aid dummies measured on an artificial head.
On the other hand, the subjective evaluation covered listeners with individual head geometries, natural head movements and a real-time hearing aid signal processing pipeline. 
Given these differences, it was unclear if the algorithm could perform well in the test condition. 
The results suggest that the GCFSnet approach is able to generalize from the training to the measurements, especially for the \emph{S0N$\pm$60 IFFM} scene with localized interfering speakers
and that head geometries and head movements are not a large issue despite being trained on static artificial scenes with only one head.
For \emph{S0Ndiff Cafeteria}, the models generalize well enough to achieve a similar performance as the traditional methods. 
The models can generalize to the simulation setup of the objective evaluation, which shows that the model can also generalize to another HRTF set (also measured on KEMAR but with a completely different loudspeaker and audio setup) and another simulation tool.
It is unclear if the relatively small model size plays an important role in this context: For a small model, it might be difficult to handle complex acoustic conditions; on the other hand, smaller models are less prone to overfitting. 
Hence, it remains to be seen if better results can be obtained by changing the size of the model, or exposing it to different HRTFs and dynamic acoustic scenes during training. 

\subsection{Support of normal-hearing listeners}
While the deep-learning enhancement algorithm analyzed in this paper was only subjectively evaluated with hearing-impaired listeners, it could be  interesting to explore its potential for supporting communication of normal-hearing listeners, e.g., by using microphone signals from consumer grade earbuds. To estimate the effect for normal-hearing listeners, we calculated HASPI scores (which can be directly interpreted as word correct scores) deactivating hearing loss compensation and HASPI configured for a normal hearing listener, averaging over SNRs as described for the main experiment. HASPI scores for localized maskers are 0.70 (unprocessed), 0.80 (\emph{MVDR}, best conventional method) and 0.92 (\emph{GCFSnet(b)}). In diffuse noise, the scores are 0.44 (unprocessed), 0.62 (\emph{MVDR}, best conventional method) and 0.72 (\emph{GCFSnet(b)}). This indicates large potential for supporting normal-hearing listeners, with average improvements of 18-45 percentage points compared to conditions unaided/gain only. These model-based predictions should be explored in future measurements with normal-hearing listeners.

\subsection{Limitations and future work}
The proposed binaural GCFSnet meets several requirements of real-world applications that were also reflected in the subjective measurements, such as providing benefits in free-field conditions and complex acoustic environments, while processing signals in real time. 
However, we also used a setup that enabled access to binaural signals without any latency, which is not compatible with wireless connections between hearing aids, something that is preferred over a wired connection by most hearing aid users. 
It is unclear how latency and lossy signal transmission would affect GCFSnet(b) and the MVDR beamformer. 
In our own related work, we have shown that the monaural GCFSnet can profit from a low-bitrate binaural link with delayed binaural features \cite{Westhausen2023} and hence we assume that a neural system with low-bitrate binaural communication would exhibit a performance similar to GCFSnet(b). 
It will be very interesting to quantify the effect of such a link in future research. 

The monaural GCFSnet does not demonstrate a significant advantage over the traditional bilateral ADM, despite its higher overall complexity. 
The overall performance might be improved when optimizing the training set, something that was not done in our study. 
Although the training data covered challenging SNRs coupled with a wide array of reverberation times, training with dynamic scenes might improve the model. 
Further, given the high variability of subjective scores, we assume that individual listeners could profit more from GCFSnet than from ADM (and the other way round), so the proposed algorithm could be added as an alternative hearing program in the future.

It would be interesting to consider other state-of-the-art DNN-based approaches with the GCFSnet, for instance (\cite{Wang2023lowlatencyIcassp} or \cite{Tokala2024}). Even if they are not compatible with hearing-aid constraints, they might provide an upper performance limit for current deep-learning systems. However, the subjective measurements performed for this study were extensive so that the number of algorithms was the upper limit for subjective measurements, so additional algorithms could not be included.

Due to the training procedure, the GCFSnet approach learned to perform noise reduction. It would therefore be very interesting to contrast its performance with established spatial filtering and additional noise reduction. However, this would at least have doubled the measurement effort, so we chose using the competitive algorithms available in the openMHA, which should foster reproducibility.

Another limitation is still relatively high latency of the complete setup which is mainly resulting from the traditional audio hardware used during the evaluation. 
This latency could be reduced by porting the algorithms directly to the PHL platform. 
Since the current PHL has only very limited resources available it was not possible to perform the required amount of optimization for the GCFSnet algorithms in the scope of this study. 
The computational cost could be lowered by implementing GCFSnet using fixed-point operations only, which could also result in an implementation that is fully compatible with a research hearing aid chip such as the smartHeap \cite{Karrenbauer2022}.

The configurations of GCFSnet evaluated in this study were trained with a fixed steering towards the front, which requires the hearing aid user to directly face the target speaker. 
While this approach effectively addresses the issue of locating the attended source, it imposes a somewhat unnatural behavioral constraint on the subjects. 
Ideally, a more adaptable steering mechanism for GCFSnet would be beneficial as suggested in \cite{Briegleb2023} and \cite{Tesch2024}. 
However, with increased flexibility comes the challenge of selecting the speaker that should be attended. 
Such a selection could be made with an external device, such as a smartphone. 
Alternatively, more sophisticated methods involving the analysis of eye gazing \cite{Kidd2013} or brain activity (invasive \cite{Han2019} or non-invasive \cite{Aroudi2020}) could be explored. 
Each of these approaches, though complex, has potential for enhancing the user experience and the functionality of GCFSnet in more dynamic and natural listening environments. 

The scope of the current setup is confined to two specific types of interferers: competing speech and diffuse cafeteria noise, each featured in a distinct scene. In real-world scenarios, individuals are likely to encounter a combination of various interferers, rather than isolated types. The rationale behind the study design was to maintain control over the experimental settings and to systematically assess performance in these two distinctly different interference situations. However, for a more comprehensive approximation to real-life conditions, future studies should consider evaluating scenarios that incorporate a combination of interferers. 

While the current study was limited to a target speaker at 0° azimuth and two symmetric, localized speech sources, we assume that the approach should generalize to asymmetric noise configurations (since the training procedure did not include any prior information about symmetry of acoustic scenes) as well as different numbers of interfering speakers (which would be in line with related studies that used DNN-based algorithms with similar training data, which were shown to generalize well to different number of competing talkers \cite{Kolbaek2017}). For a larger number of speakers, the noise field would become diffuse and become very similar to one of the acoustic conditions explored above. Finally, since our system is trained to enhance speech from the front, competing speakers at different angles should be effectively suppressed.

\section{Conclusion}
This study explored the potential of deep neural networks for speech enhancement in hearing aids, based on a recurrent network with low latency and a small computational footprint, the GCFSnet.

Speech intelligibility for hearing-impaired listeners is clearly improved with the GCFSnet compared to using unprocessed signals.

In the presence of localized interferers, deep enhancement based on binaural communication produces the highest intelligibility compared to established hearing aid signal enhancement strategies considered here.

For data pooled over hearing-impaired listeners, the objective metrics HASPI and MSBG+MBSTOI are good predictors for performance differences between signal enhancement strategies.

\section*{Acknowledgments}
This work was supported by the Deutsche Forschungsgemeinschaft (DFG, German Research Foundation) under Germany’s Excellence Strategy – EXC 2177/1 - under Grants 390895286 and 352015383 - SFB 1330 C6.
We would like to thank Mathias Blau and Felix Stärz at the Jade Hochschule in Oldenburg for providing their help and expertise and the measurement setup for the high resolution HRTF set.
We also would like to thank Annika Meyer-Hilberg and Janique Reinwaldt for conducting the listening experiments.

\section{References Section}
\bibliographystyle{IEEEtran}
\bibliography{westhausen_24}

\newpage

\vfill

\end{document}